\documentclass[a4paper,fleqn,usenatbib]{mn2e}

\usepackage[T1]{fontenc}
\usepackage{ae,aecompl}
\usepackage{color}

\usepackage{graphicx}    
\usepackage{amsmath}    
\usepackage{amssymb}    
\usepackage{upgreek}
\def\beq{\begin{equation}}
\def\eeq{\end{equation}}
\def\bea{\begin{eqnarray}}
\def\eea{\end{eqnarray}}
\def\be{\begin{equation}}
\def\ee{\end{equation}}

\def\bse{\begin{subequations}}
	\def\ese{\end{subequations}}

\usepackage{calrsfs}

\def\apj{{ApJ}}%
\def\apjl{{ApJL}}%
\def\apss{{Ap\&SS}}
\def\aap{{A\&A}}%
\def\mnras{{MNRAS}}%
\def\na{{New Astron.}}%
\def\pasj{{PASJ}}%
\def\ssr{{Space Science Reviews}}%

\title[Shocks in relativistic viscous accretion flows]{Shocks in relativistic viscous accretion flows around Kerr black holes}

\author[Dihingia et al]{
Indu K. Dihingia$^{1}$\thanks{E-mail: i.dihingia@iitg.ac.in},
Santabrata Das$^{1}$\thanks{E-mail: sbdas@iitg.ac.in},
Debaprasad Maity$^{1}$\thanks{E-mail: debu@iitg.ac.in} and
Anuj Nandi$^{2}$\thanks{E-mail: anuj@ursc.gov.in}
\\
$^{1}$Indian Institute of Technology Guwahati, Guwahati, 781039, Assam, India\\
$^{2}$Space Astronomy Group, ISITE Campus, U. R. Rao Satellite Center, Outer Ring Road, Marathahalli, Bangalore, 560037, India
}

\date{Accepted XXX. Received YYY; in original form ZZZ}

\pubyear{2016}

\begin{document}
\label{firstpage}
\pagerange{\pageref{firstpage}--\pageref{lastpage}}
\maketitle

\begin{abstract}
	
		We study the relativistic viscous accretion flows around the Kerr black holes. We present the governing equations that describe the steady state flow motion in full general relativity and solve them in 1.5D to obtain the complete set of global transonic solutions in terms of the flow parameters, namely specific energy (${\cal E}$), specific angular momentum (${\cal L}$) and viscosity ($\alpha$). We obtain a new type of accretion solution which was not reported earlier. Further, we show for the first time to the best of our knowledge that viscous accretion solutions may contain shock waves particularly when flow 	simultaneously passes through both inner critical point ($r_{\rm in}$) and outer critical point ($r_{\rm out}$) before entering into the Kerr black holes. We examine the shock properties, namely shock location ($r_s$) and compression ratio ($R$, the measure of density compression across the shock front) and show that shock can form for a large region of parameter space in ${\cal L}-{\cal E}$ plane. We study the effect of viscous dissipation on the shock parameter space and find that parameter space shrinks as $\alpha$ is increased. We also calculate the critical viscosity parameter ($\alpha^{\rm cri}$) beyond which standing shock solutions disappear and examine the correlation between the black hole spin ($a_k$) and $\alpha^{\rm cri}$. Finally, the relevance of our work is conferred where, using $r_s$ and $R$, we 	empirically estimate the oscillation frequency of the shock front ($\nu_{QPO}$) when it exhibits Quasi-periodic (QP) variations. The obtained results indicate that the present formalism seems to be potentially viable to account for the QPO frequency in the range starting from milli-Hz to kilo-Hz as $0.386~{\rm Hz}\le \nu_{QPO} \left(\frac{10M_\odot}{M_{BH}} \right) \le 1312$ Hz for $a_k=0.99$, where $M_{BH}$ stands for the black hole mass.
\end{abstract}


\section{Introduction}

The accretion process onto a black hole is believed to be the most efficient energy release mechanism because of the fact that it is an order of magnitude stronger than the nuclear fusion reactions \citep{Frank-etal2002}. As black holes themselves do not emit any radiation, therefore, one compels to rely on the study of accretion flows
in order to understand the astrophysical black holes.
And, because of the strong gravity of the black holes, it is necessary to examine the relativistic accretion flows considering the general relativistic framework. 

In the early seventies, a comprehensive study of relativistic accretion flow around the Kerr black holes was carried out by \cite{Novikov-Thorne1973}. Latter, \cite{Fukue1987} examined the transonic properties of the inflowing matter considering full relativistic treatment. \cite{Riffert_Herold1995} reported the correct description of the thin accretion disc structure around the Kerr black holes. Meanwhile, \cite{Chakrabarti1996a,Chakrabarti1996b} examined the relativistic accretion flow assuming weak viscosity limit and \cite{Peitz_Appl1997} studied the relativistic viscous accretion flow considering the polytropic equation of state. Subsequently, \cite{Gammie1998,Popham1998} pointed out the importance of relativistic equation of state (EoS) while studying the relativistic accretion flow. \cite{Chattopadhyay-Chakrabarti2011} investigated the effect of fluid composition on the accretion flow properties around the Schwarzschild black holes. Recently, \cite{Chattopadhyay_Kumar2016} studied the accretion-ejection solutions in full general relativity considering non-rotating black holes\textcolor{red}{,} and they extended the work further for rotating black holes as well \cite[]{Kumar-Chattopadhyay2017}.

It may be noted that some of the above studies examined the phenomena of shock waves where accretion flow variables encounter discontinuous transitions \citep{Fukue1987,Chakrabarti1996a,Chakrabarti1996b,Chattopadhyay-Chakrabarti2011,Chattopadhyay_Kumar2016,Kumar-Chattopadhyay2017}. In reality, there are various astrophysical phenomena associating the shock waves, namely supernova explosions, various outburst phenomena, shocks in astrophysical jets and winds \citep{Fukue2019}. This truly indicates that shocks are common in astrophysical environments.
However, the study of shock waves in the relativistic viscous accretion flow around the Kerr black holes remains unexplored till date. In this context, questions naturally arise whether shocks continue to be present or not under strong gravity? 
If so, what would be the influence of the relativistic equation of state (EoS) on the shock properties? How is the shock location affected due to the viscosity as well as the black hole spin? Can the modulation of shock front render the Quasi-periodic Oscillation (QPO) phenomena commonly observed in Galactic black hole sources? In the present work, for the first time to our knowledge, we intend to answer these questions.

In an accretion process, rotating matter begins its journey
towards the black hole with negligible radial velocity from the outer edge of the disc. As the flow moves inward, the radial velocity gradually increases and 
eventually, subsonic flow changes its sonic state at the critical point to become supersonic before crossing the event horizon. Depending on the input parameters, the flow may contain multiple critical points,  and in general, the flow of this kind first becomes supersonic much before the horizon \cite[]{Fukue1987,Chakrabarti1989,Das-etal2001a}. In this scenario, supersonic matter experiences a centrifugal barrier that causes the piling of matter around the black hole. Eventually, such barrier triggers the discontinuous transition of the flow variables in the form of shock waves provided the relativistic shock conditions are satisfied \cite[]{Taub1948}. After the shock transition, flow momentarily slows down, however, gradually picks its radial velocity as it moves inward and ultimately enters into the black hole after crossing another critical point usually located close to the horizon. This renders the complete shock induced global accretion solution and solutions of this kind have been examined by several researchers \cite[]{Fukue1987,Chakrabarti1989,Yang_Kafatos1995,Lu-etal99,Becker_Kazanas2001,Das-etal2001a,Fukumura-Tusuruta04,Chakrabarti-Das2004,Das2007,Sarkar_Das2016,Dihingia_etal2018b,Dihingia_etal2018a,Dihingia-etal2019}. Due to shock compression, post-shock flow becomes hot and dense that results a puffed up torus like structure surrounding the black hole which is equivalently called as post-shock corona (hereafter PSC) \citep{Aktar_etal2015}. Interestingly, because of the extra thermal energy exists across the shock front, a part of the inflowing matter is deflected at PSC to produce the precursor of the bipolar jets along the rotation axis of the disc \citep{Chakrabarti1999,Das-etal2001b,Chattopadhyay-Das2007,Das-Chattopadhyay2008,Aktar_etal2015,Aktar_etal2017,Aktar-etal2019}. These findings are also confirmed by the numerical simulations \citep{Molteni_etal1996,Lanzafame-etal1998,Das_etal2014,Okuda-Das2015,Lee-etal2016}. 

Incidentally, the same PSC seems to be responsible for the emission of hard radiations observed from the active galactic nuclei (AGNs) and Galactic black hole (GBH) sources. Usually, soft photons from the pre-shock disc are intercepted at the PSC and reprocessed after interacting with the swarm of hot electrons to generate high energy photons via inverse Comptonization mechanism \citep{Chakrabarti_Titarchuk1995,Mandal2005,Mandal-Chakrabarti2008,Iyer_etal2015}. Moreover, \cite{Molteni_etal1996} showed through numerical simulation that when infall time scale matches with the cooling time scale of the accreting matter, resonance oscillation of PSC takes place. As PSC modulates, emergent hard radiations also exhibit non-steady variations which are in general quasi-periodic in nature \citep{Lee_etal2011,Das_etal2014,Sukova_Janiuk2015,Sukova_etal2017,Okuda-etal2019}. Hence, the modulation of PSC perhaps be potentially viable to account for the QPO phenomena commonly observed in Galactic Black hole sources \citep{Belloni_etal2002,Homan-Belloni2005,Remillard_McClintock2006,Nandi_etal2012,Iyer_etal2015,Nandi-etal2018}. In addition, episodic ejections of the matter are also seen as a consequence of PSC undulations \citep{Das_etal2014}. Overall, all the above findings generally supplement the importance of PSC as its role  
appears to be very much appealing in order to explain the astrophysical sources harboring black holes. 

Being motivated with this, in this work, we study relativistic, viscous, advective, accretion flow around a Kerr black hole. Although 3D time-dependent modeling of general relativistic flow exists in the literature, in this work, we consider steady state 1.5D flow structure in order to obtain the analytical accretion solutions.
Here, we adopt the relativistic hydrodynamic framework to study the flow dynamics \citep{Rezzolla-Zanotti2013}. 
In addition, we consider the relativistic equation of state to describe the  accreting plasma  \citep{Chandrasekhar1939,Synge1957,Cox_Giuli1968}. 
Incorporating all these, we carry out the critical point analysis and obtain all possible global transonic accretion solutions around the Kerr black holes. Further, we employ the relativistic shock conditions \citep{Taub1948} and calculate the shock induced global accretion solutions. We study the shock properties, namely shock location ($r_s$) and compression ratio ($R$, measure of density compression across the shock front) in terms of the input parameters ($i.e.$, specific energy (${\cal E}$) and specific angular momentum (${\cal L}$) and examine the role of viscosity ($\alpha$) and black hole spin ($a_k$) in deciding the flow characteristics. Moreover, we identify the parameter space in ${\cal L}- {\cal E}$ plane that admits shock and finds that shock parameter space is shrunk with the increase of viscous dissipation. We obtain the critical viscosity parameter ($\alpha^{\rm cri}$) beyond which standing shock solutions disappear and investigate $a_k - \alpha^{\rm cri}$ correlation for shocks. Since $r_s$ eventually measures the size of PSC, we phenomenologically calculate the QPO frequency of the PSC modulation ($\nu_{QPO}$) which is equivalent to the inverse of the infall time scale of post-shock matter. We find that $\nu_{QPO}$ lies in the range of $0.386-1312$ Hz 
for $M_{\rm BH}=10M_\odot$ and $a_k=0.99$ which seems to be fairly consistent with observation \cite[]{Remillard_McClintock2006,Belloni-etal2012,Nandi_etal2012,Belloni-Stella2014,Iyer_etal2015,Motta2016,Sreehari-etal2019a,Sreehari-etal2019b}.

We organize the paper as follows. In \S2, we present the relativistic hydrodynamics in Kerr space time and in \S3, we discuss the modeling of the accretion flow. In \S4, we carry out the critical point analysis and present the global transonic solutions. In \S5, we discuss the global solutions with shock, shock properties, shock parameter space, and shock mediated QPOs. Finally, in \S6, we present the concluding remarks.

\section{Relativistic Hydrodynamics in Kerr spacetime} 

We study the hydrodynamics of accretion flow in a generic stationary axisymmetric space-time. Here, there exist two mutually computing killing vectors along $(t,\phi)$ directions. The remaining space-like coordinates are $(r,\theta)$ which are mutually orthogonal and also orthogonal to the two killing vectors at every point in space-time. With this coordinate system, a stationary axisymmetric space-time is written as,
 
$$
ds^2 = g_{\mu\nu} dx^\mu dx^\nu
\eqno(1)
$$$$
~= g_{tt}dt^2 + 2g_{t\phi}dtd\phi+ g_{\phi\phi} d\phi^2 + 
g_{rr} dr^2 + g_{\theta\theta} d\theta^2 ,
$$
where the indices $\mu$ and $\nu$ run from $0$ to $3$ representing $t$, $r$, $\theta$ and $\phi$ coordinates, respectively. Due to the presence of the two killing vectors ($l^\mu_t=\partial_t, l^\mu_\phi = \partial_\phi$), the metric coefficients are in general expressed as the functions of coordinates $(r,\theta)$. The non-zero metric elements in Boyer-Lindquist coordinates are given by \citep{Boyer_Lindquist1967},
$$
g_{tt}=-\left(1-\frac{2 r}{\Sigma}\right),~~~~~~ g_{t\phi}=-\frac{2 a_{\rm k} r \sin ^2\theta }{\Sigma},
$$$$
g_{rr}=\frac{\Sigma }{\Delta},~~~ g_{\theta\theta}=\Sigma, ~~{\rm and}~~ 
g_{\phi\phi}=\frac{A\sin ^2\theta}{\Sigma},
$$
where $\Sigma=a_{\rm k}^2 \cos ^2\theta+r^2$, $\Delta=a_{\rm k}^2+r^2-2 r$ and 
$A=\left(a_{\rm k}^2+r^2\right)^2-a_{\rm k}^2\Delta \sin^2\theta $, respectively. Here, the specific spin of the black hole is written as $a_{\rm k}=J/M_{\rm BH}$, where $M_{\rm BH}$ denotes the mass of the black hole. To express the physical quantities, we use a convenient 
unit system as $G=M_{\rm BH}=c=1$, where $G$ is the gravitational constant, $c$
is the speed of light. In this system, length, time and angular momentum are expressed in unit of $GM_{\rm BH}/c^2$,  $GM_{\rm BH}/c^3$ and $GM_{\rm BH}/c$, respectively.

The relativistic hydrodynamics is governed by the conservation energy momentum 
and particle number as,
$$
T^{\mu\nu}_{;\nu}=0,\qquad (\rho u^\nu)_{;\nu}=0,
\eqno(2)
$$
where, $T^{\mu\nu}$ denotes the energy momentum tensor, $\rho$ is the density of the
flow, and $u^\nu$ are the components of four velocities supplemented with the constraint $u^\mu u_\mu=-1$. The energy momentum tensor is written as,
$$
T^{\mu\nu} = (e+p)u^\mu u^\nu + pg^{\mu\nu} + \pi^{\mu\nu},
\eqno(3)
$$
where $e$ and $p$ are the local energy density and local isotropic pressure of the flow. The last term in the right hand side of equation (3) represents the viscous stress tensor. By presuming the shear that gives rise to the viscosity, we have $\pi^{\mu\nu}=-2\eta\sigma^{\mu\nu}$, where $\eta$ is the viscosity coefficient and the shear tensor is given by \citep{Peitz_Appl1997},
$$
\sigma_{\mu\nu}=\frac{1}{2}\left[(u_{\mu;\nu}+ u_{\nu;\mu}+a_\mu u_\nu + 
a_\nu u_\mu) - \frac{2}{3}\zeta_{exp}h_{\mu\nu}\right],
\eqno(4)
$$  
where $a_\mu~(= u_{\mu;\gamma}u^\gamma)$ is the four acceleration, 
$\zeta_{exp}~(=u^\gamma_{;\gamma})$ is expansion of the fluid world line 
and $h_{\mu\nu}~(= g_{\mu\nu}+u_\mu u_\nu)$ is the projection tensor. Here,
$\eta=\rho\nu$, where $\nu$ is the kinetic viscosity. 

By projecting the energy momentum conservation equation along the $i$-th direction, we obtain the Navier-Stokes equation as,
$$
h^i_\mu T^{\mu\nu}_{;\nu}=
(e+p)u^\nu u^i_{;\nu} + (g^{i\nu} + u^i u^\nu)p_{,\nu} 
+ h^i_\mu\pi^{\mu\nu}_{;\nu}=0, 
\eqno(5)
$$
where $i =1,2,3$ and $h^i_{\mu}u^{\mu}=0$.
Similarly, the energy generation equation ($i.e.$, first law of thermodynamics) is given by $u_\mu T^{\mu\nu}_{;\nu}=0$ which takes the form, 
$$
u^\mu\bigg[\left(\frac{e+p}{\rho}\right)\rho_{,\mu} 
- e_{,\mu}\bigg]+u_\mu \pi^{\mu\nu}_{;\nu} =0,
\eqno(6)
$$
where the last term in the left hand side of equation (6) represents the viscous heating term and the specific enthalpy of the flow is given by $h=(e+p)/\rho$. To avoid complexity, here we ignore the radiative cooling processes.

In order to solve the hydrodynamical equations that govern the accretion flow around black holes, we require an exact relation among $e$, $\rho$ and $p$, which is commonly known as equation of state (EoS). For relativistic fluid, we consider an EoS \citep{Chattopadhyay2009} which is given by,
$$
e = n_em_ef=\frac{\rho}{\tau}f,
\eqno(7)
$$
where
$$
f = (2-\xi)\bigg[1 + \Theta\left(\frac{9\Theta + 3}{3\Theta + 2}\right)\bigg] +
\xi\bigg[ \frac{1}{\chi} + \Theta\left(\frac{9\Theta + 3/\chi}
{3\Theta + 2/\chi}\right)\bigg].
\eqno(8)
$$
Here, $n_e~(n_p)$ and $m_e~(m_p)$ denote the number density and mass of the electron (ion) and $\Theta=k_{\rm B}T/m_ec^2$ is the dimensionless temperature of the flow. In addition, $\tau=[2-\xi(1-1/\chi)]$, where we use $\xi = n_p/n_e$ and $\chi = m_e/m_p$, respectively. In this work, we use $\xi=1$ all throughout unless stated otherwise.
With this, the polytropic index $(N)$, the ratio of specific heats $(\Gamma)$ and the sound 
speed $(a_s)$ are defined as,
$$
N = \frac{1}{2}\frac{df}{d\Theta}; \quad \Gamma = 1 + \frac{1}{N}; \quad{\rm and}
\quad a_s^2 = \frac{\Gamma p}{e+p} = \frac{2\Gamma\Theta}{f + 2\Theta}.
\eqno(9)
$$

\section{Modeling of Accretion flow}

We consider a steady, viscous, advective accretion disc confined around the black hole equatorial plane. Hence, for simplicity, we assume $\theta = \pi/2$ and $v_\theta \sim 0$ throughout the study. In addition, we define the angular velocity $v^2_\phi = (u^\phi u_\phi)/(-u^t u_t)$ and the associated bulk azimuthal Lorentz factor as $\gamma^2_\phi =1/(1-v^2_\phi)$. Similarly, we also define the radial three velocity in the co-rotating frame as $v^2=\gamma^2_\phi v^2_r$, where $v^2_r=(u^r u_r)/(-u^t u_t)$ and the associated bulk Lorentz factor $\gamma^2_v =1/(1-v^2)$. Employing these definitions of velocities, we rewrite the second part of the equation (2) in the integrated form as, 
$$
\dot{M}=-4\pi v \gamma_v \rho H \sqrt{\Delta},
\eqno(10)
$$
where $\dot{M}$ is accretion rate which we treat as global constant and $H$ is the local half-thickness of the disc. Considering the thin disc approximation, we calculate the functional form of $H$ as \citep{Riffert_Herold1995,Peitz_Appl1997}, 
$$
H^2 = \frac{pr^3}{\rho \mathcal{F}},\qquad
\mathcal{F}=\gamma_\phi^2\frac{(r^2 + a_{\rm k}^2)^2 + 2\Delta a_{\rm k}^2}
{(r^2 + a_{\rm k}^2)^2 - 2\Delta a_{\rm k}^2}.
\eqno(11)
$$

In the next, we obtain the radial momentum equation in the co-rotating frame by setting $i=r$ and is given by, 
$$
v\gamma_v^2\frac{dv}{dr} + \frac{1}{e+p}\frac{dp}{dr} + \left(\frac{\partial \Phi}{\partial r}\right)_{\lambda}=0,
\eqno(12)
$$
where 
$$
\Phi =\frac{1}{2}\ln \left[ \frac{r\Delta}{a_{\rm k}^2 (r+2)-4 a_{\rm k} \lambda +r^3-\lambda ^2 (r-2)}\right],
$$
and $\lambda =-u_\phi/u_t$.
In equation (12), following \cite{Gammie1998,Popham1998}, we neglect the viscous
acceleration term in the radial momentum equation.

Employing the killing vectors $l^\mu_t$ and $l^\mu_\phi$, we obtain two conserved quantities as,
$$
{\cal E}=-\left(hu_t - \frac{2\nu\sigma^r_t}{u^r}\right)
\eqno(13)
$$
and
$$
{\cal L}=hu_\phi - \frac{2\nu\sigma^r_\phi}{u^r},
\eqno(14)
$$
where, $\nu = \alpha a_s H$, $\alpha$ being the viscosity parameter \cite[]{Gammie1998,Popham1998}. Since the terms involved in equations (13) and (14) have the dimensions of energy per unit mass and angular momentum, we call ${\cal E}$ and ${\cal L}$ as global specific energy and bulk specific angular momentum of the flow, respectively.

Subsequently, we calculate $\sigma^r_\phi$ and $\sigma^r_t$ from equation (4), which are given by,
$$
2\sigma^r_\phi=u^r_{;\phi} + g^{rr}u_{\phi;r} + a^r u_\phi + a_\phi u^r - \frac{2}{3}\zeta_{exp}u^ru_\phi,
\eqno(15)
$$
and
$$
2\sigma^r_t=u^r_{;t} + g^{rr}u_{t;r} + a^r u_t + a_t u^r - \frac{2}{3}\zeta_{exp}u^ru_t.
\eqno(16)
$$

Using the velocity definitions, we re-write equation (15) and equation (16) in the following forms as,
$$
2\sigma^r_\phi= {\cal A}_1 + {\cal A}_2 \frac{dv}{dr} + {\cal A}_3 \frac{d\lambda}{dr}
\eqno(17)
$$
and
$$
2\sigma^r_t= {\cal B}_1 + {\cal B}_2 \frac{dv}{dr} + {\cal B}_3 \frac{d\lambda}{dr},
\eqno(18)
$$
where the coefficients ${\cal A}_1$, ${\cal A}_2$, ${\cal A}_3$, ${\cal B}_1$, ${\cal B}_2$ and ${\cal B}_3$ are the functions of the flow variables and their functional forms are given in Appendix-A. Since the first order derivatives of $v$ and $\lambda$ in the shear tensor yields the governing equations of the relativistic flow as second order, it is difficult to solve them. Therefore, to avoid complexity, we neglect the terms containing the higher order derivatives of $v$ and $\lambda$ in equation (17) and (18) and with these approximations, we are left with $2\sigma^r_\phi = {\cal A}_1$ and $2\sigma^r_t = {\cal B}_1$, respectively.

Since ${\cal E}$ and ${\cal L}$ are conserved quantities, there derivatives vanishes and accordingly, we have, 
$$
\frac{d{\cal E}}{dr}={\cal E}_0 + {\cal E}_1 \frac{dv}{dr} + {\cal E}_2 \frac{d\Theta}{dr} + {\cal E}_3 \frac{d\lambda}{dr} = 0 
\eqno(19)
$$
and 
$$
\frac{d{\cal L}}{dr}={\cal L}_0 + {\cal L}_1 \frac{dv}{dr} + {\cal L}_2 \frac{d\Theta}{dr} + {\cal L}_3 \frac{d\lambda}{dr} = 0,
\eqno(20)
$$
where ${\cal E}_0$, ${\cal E}_1$, ${\cal E}_2$, ${\cal E}_3$, ${\cal L}_0$, ${\cal L}_1$, ${\cal L}_2$ and ${\cal L}_3$ are functions of the flow variables and their expression are given in Appendix-B.

We simultaneously solve equations (10), (12), (19) and (20) to obtain the wind equation as,
$$
\frac{dv}{dr}=\frac{N(r,v,\lambda, \Theta)}{D(r,v,\lambda, \Theta)},
\eqno(21)
$$
where $N(r,v,\lambda, \Theta)$ and $D(r,v,\lambda, \Theta)$ are the functions of the flow variables and their algebraic expressions are provided in the Appendix-C. Further, the gradients of $\Theta$ and $\lambda$ is obtained respectively as,
$$
\frac{d\Theta}{dr}=\frac{\left({\cal L}_3 {\cal E}_1-{\cal L}_1 {\cal E}_3\right)}
{{\cal L}_2 {\cal E}_3-{\cal L}_3 {\cal E}_2}\frac{dv}{dr}
+\frac{{\cal L}_3 {\cal E}_0-{\cal L}_0 {\cal E}_3}{{\cal L}_2 {\cal E}_3-{\cal L}_3 {\cal E}_2};
\eqno(22) 
$$
and
$$
\frac{d\lambda}{dx}=\frac{\left({\cal L}_2 {\cal E}_1-{\cal L}_1 {\cal E}_2\right)}
{{\cal L}_3 {\cal E}_2-{\cal L}_2 {\cal E}_3}\frac{dv}{dr}
+\frac{{\cal L}_2 {\cal E}_0-{\cal L}_0 {\cal E}_2}{{\cal L}_3 {\cal E}_2-{\cal L}_2 {\cal E}_3},
\eqno(23)
$$
where ${\cal E}_{i}$ and ${\cal L}_{i}$ with $i=0, 1, 2, 3$ are described in Appendix-B.

\section{Critical point analysis and Global solutions}
\begin{figure}
	\includegraphics[width=0.485\textwidth]{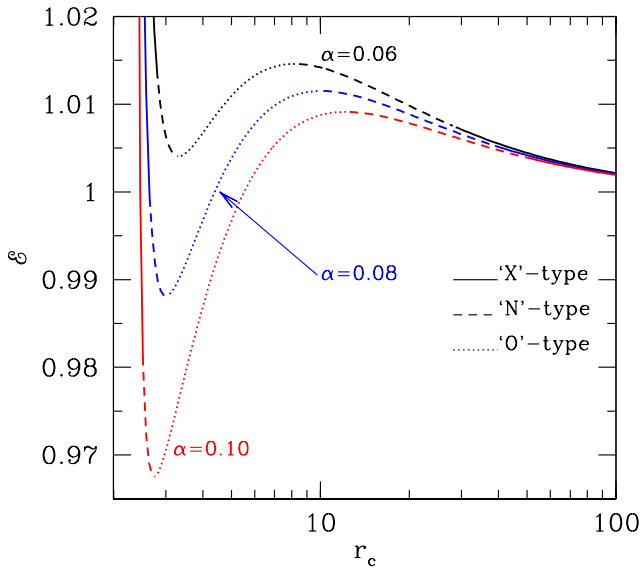}
	\caption{Plots of global specific energy (${\cal E}$) as a function of critical point locations ($r_{\rm c}$) for three different viscosity parameters ($\alpha$) marked on the figure. Here. we consider ${\cal L} = 1.80$ and $a_k=0.99$, respectively. Solid, dotted and dashed curves represent the results corresponding to saddle (X), nodal (N) and spiral (O) type critical points. See text for details. 
	}
\end{figure}

We carry out the critical point analysis following the standard procedure \cite[and references therein]{Dihingia_etal2018c} and examine the properties of the critical points in terms of the input parameters of the flow. At the critical point, equation (21) takes $dv/dr=0/0$ form, where the conditions $N(r,v,\lambda, \Theta)=0$ and $D(r,v,\lambda, \Theta)=0$ are known as the critical point conditions.
Applying the l$^{\prime}$Hospital rule, we calculate the radial velocity gradient $(dv/dr)_{\rm c}$ at the critical points ($r_{\rm c}$).  
Depending on the values of $(dv/dr)_{\rm c}$, the nature of the critical points are classified. In reality, $(dv/dr)_{\rm c}$ usually possesses two values. When the values of $(dv/dr)_{\rm c}$ are real and of opposite sign, the critical point is called as saddle type (hereafter `X-type'). For real and same sign of $(dv/dr)_{\rm c}$ values yield nodal type critical point (hereafter `N-type'). When both values of $(dv/dr)_{\rm c}$ are imaginary, the nature of the critical point becomes spiral type (hereafter `O-type'). Based on the above classifications, we examine how the different types of critical points spread along the radial direction. The obtained results are depicted in Fig. 1, where we plot the variation of global specific energy (${\cal E}$) as a function of critical points ($r_c$) for three different viscosity parameters as $\alpha=0.06, 0.08$ and $0.10$, respectively. Here, we choose $a_{\rm k}=0.99$ and ${\cal L}=1.80$. In the figure, viscosity parameters are marked, and solid, dashed and dotted curves denote the `X-type', `N-type' and `O-type' critical points, respectively. We observe that for a given $\alpha$, different types of critical points are located along the increasing radial coordinate following the sequence of saddle --- nodal --- spiral --- nodal --- saddle types, respectively. For a given ${\cal E}$, the flow may have maximum three critical points, out of which one is `O-type' and the other two may be either `X-type' or `N-type' or their combinations depending on the input parameters. Moreover, we find that as the viscosity of the flow is increased, a part of the `X-type' critical points from both inner and outer regions are gradually replaced by the `N-type' critical points. In reality, `X-type' critical points are specially important as the accretion flow around the black holes can only pass through it. In addition, it may be noted that the accretion flows passing through the `N-type' critical points are found to be unstable \cite[]{Kato_etal1993}. Moreover, when the critical point resides near the horizon, it is called as inner critical point ($r_{\rm in}$) whereas when it forms far away from the black hole, it is called as outer critical point ($r_{\rm out}$). And, when accretion flow possesses multiple `X-type' critical points, it may experience discontinuous shock transition in between the inner and outer critical points, provided the relativistic shock conditions are satisfied \cite[]{Taub1948}. Accretion solutions of this kind are very much important as they have the potential to explain the observational findings \cite[and references therein]{Chakrabarti_Titarchuk1995,Wu-etal2002,Nandi_etal2012,Iyer_etal2015,Sukova_Janiuk2015,Fukumura-etal2016,Sukova_etal2017} and therefore, in the subsequent section, we investigate the global shock solutions around the Kerr black holes.

\begin{figure}
	\includegraphics[width=0.485\textwidth]{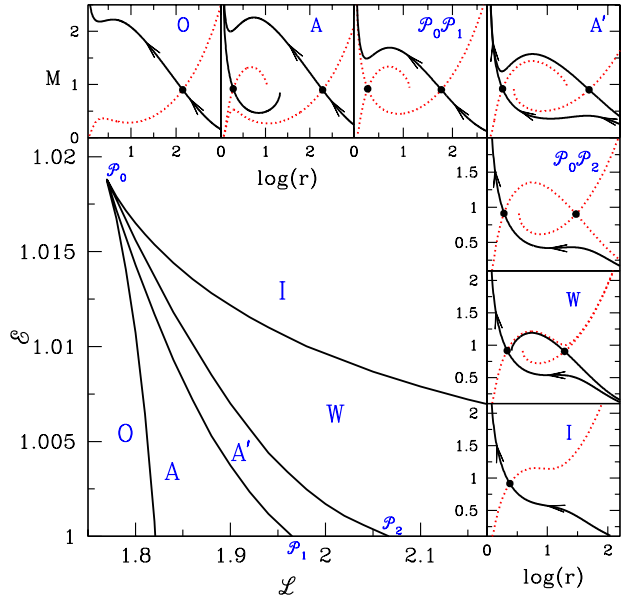}
	\caption{Sub-division of ${\cal L}-{\cal E}$ parameter space on the basis of the type of accretion solutions. Different regions are marked on the figure as `O', `A', `${\cal P}_0{\cal P}_1$', `A$^\prime$', `${\cal P}_0{\cal P}_2$', `W' and `I', and the corresponding Mach number ($M=v/a_s$) plots are depicted on the boxes. The filled circle in each boxes represents the critical point locations of the flow and the arrow indicates the overall direction of the flow motion. Here, we choose $a_k=0.99$ and $\alpha=0.05$. See text for details.
		}
\end{figure}

To obtain the accretion solution, one requires to solve the equations (21-23) simultaneously. Since the accretion flow around the black hole is necessarily transonic, it is advantageous to start integrating equations (21-23) from the critical point itself. Hence, we choose a set of input parameters as (${\cal E}, {\cal L}, \alpha, a_k$) and employing the critical point conditions, we solve equations (13) and (14) to calculate the radial velocity ($v_c$), temperature ($\Theta_c$) and angular momentum ($\lambda_c$) at the critical point ($r_c$). Using these flow variables, we integrate equations (21-23) once inwards from the critical point up to the horizon and then outward up to a large distance equivalent to the outer edge of the disk ($r_{\rm edge}$). Finally, we join these two parts of the solution to obtain a complete global accretion solution around the black holes. It may be noted that solving equations (21-23) using the boundary values supplied at $r_{\rm edge}$ yields the identical accretion solution as described above.

Next, we investigate the general behavior of the accretion solutions, and for that, we subdivide the ${\cal L}-{\cal E}$ parameter space according to the nature of the accretion solutions. We find five different types of physically acceptable accretion solutions and therefore, we sub-divide the parameter space in five regions marked as O, A, A$^\prime$, W and I in Fig. 2. Here, we choose $a_k=0.99$ and $\alpha = 0.05$.
Typical solutions obtained from these five regions are depicted in the inset boxes which are marked. In each box, Mach number ($M$) of the flow is plotted as function of the logarithmic radial coordinate ($r$) where solid curve represents the accretion branch while the dashed curve denotes the wind branch.
We find that depending on the input parameters, the flow may contain single or multiple critical points which are shown using the filled circles. In addition, arrows indicate the overall direction of flow motion towards the black hole. {\it Here, we find a new type of  accretion solution} (A$^\prime$) {\it where inflowing matter having identical (${\cal L}, {\cal E}$) has the option to pass through either outer or inner critical points}. In order to resolve the degeneracy of the accretion solution, we calculate the entropy of the flow just outside the horizon and find that solution passing through the inner critical point has high entropy content. Since nature favors the high entropy flow \cite[]{Becker_Kazanas2001}, solution passing through the inner critical point is physically acceptable. Example of this kind of solution is shown in the box marked A$^\prime$. Interestingly, supersonic flow after crossing the outer critical point has the possibility to join with the sub-sonic branch via shock transition (see \S 5), however, we point out that relativistic shock conditions (see equation (24) below) are not satisfied for this type of accretion solutions.
We also find accretion solutions with a special topological property where a single integral curve passes through both critical points simultaneously. This type of solutions are found along the line ${\cal P}_o-{\cal P}_1$ and ${\cal P}_o-{\cal P}_2$ in the ${\mathcal L}-{\cal E}$ parameter space and illustrated in the boxes marked ${\cal P}_o{\cal P}_1$ and ${\cal P}_o{\cal P}_2$, respectively. It is noteworthy to mention that accretion solutions depicted in panels marked `W' and `I' are identical to the advection dominated accretion flow (ADAF) solutions \citep[and references therein]{Narayan-etal97}.

\section{Accretion solution with shock}

In this section, we study the properties of the 
accretion flow that possesses multiple critical points. In reality, accretion flow begins its journey from the outer edge of the disk ($r_{\rm edge}$) with negligible radial velocity ($v \ll c$) and gradually gains its radial velocity as it accretes towards the black hole. At the outer critical point ($r_{\rm out}$), flow experiences smooth sonic state transition from subsonic to the supersonic regime and continues to proceed further. Meanwhile, centrifugal repulsion starts to become profound, and it plays a preponderant role against gravity to slow down the inflowing matter. Because of this, accreting matter piles up in the vicinity of the black hole, and a centrifugal barrier is developed. 
This process continues unless the centrifugal barrier triggers the discontinuous transitions of the flow variables in the form of a shock wave. Due to the shock transition, the supersonic flow jumps into the subsonic branch and eventually picks up its radial velocity while moving further inwards. Ultimately, accretion flow again becomes supersonic after passing through the inner critical point ($r_{\rm in}$) before falling into the black hole.

\begin{figure}
	\includegraphics[width=0.485\textwidth]{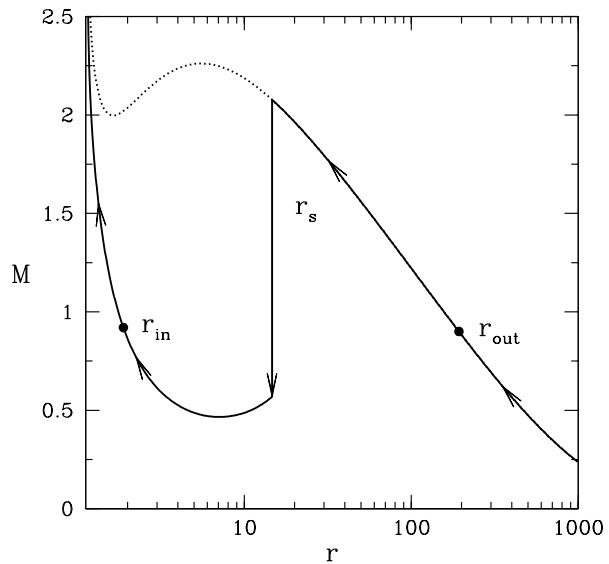}
	\caption{Illustration of a typical shock induced global accretion solution around a Kerr black hole. Here, input parameters are chosen as ${\cal L}=1.90$, ${\cal E}=1.001$, $\alpha = 0.05$ and $a_k=0.99$, respectively. Inner and outer critical points are calculated as $r_{\rm in} = 1.8898$ and $r_{\rm out}=192.9923$. Flow encounters shock transition at $r_s=14.67$ indicated by the vertical arrow. See text for details.
	}
\end{figure}

To illustrate the above scenario, we depict a shock induced global accretion solution around a black hole in Fig. 3, where Mach number ($M$) of the flow is plotted as function of radial coordinate ($r$). Here, we choose the input parameters as
${\cal E} = 1.001$, ${\cal L}=1.90$, $\alpha=0.05$ and $a_k=0.99$, respectively and find the inner and outer critical points as $r_{\rm in} = 1.8898$ and $r_{\rm out} = 192.9923$. After crossing $r_{\rm out}$, the flow has the possibility to enter into the black hole supersonically as shown by the dotted curve. However, flow experiences shock transition in between $r_{\rm in}$ and $r_{\rm out}$ as the relativistic shock conditions are favorable. In reality, the shock solution is preferred over the shock-free solution because of its high entropy content \cite[]{Becker_Kazanas2001}.
We calculate the location of the shock radius using the relativistic shock conditions \citep{Taub1948}, which are given by,  
$$\begin{aligned}
&[\rho u^r]=0, \qquad [(e+p)u^tu^r]=0,\\
&{\rm and}\quad[(e+p)u^ru^r + pg^{rr}]=0,\\
\end{aligned}\eqno(24)$$
where we assume the shock to be thin and the quantities within the square bracket represent the difference of their values across the shock front. Here, shock location is calculated as $r_s=14.67$ which is indicated by the vertical arrow and the overall direction of the flow motion is indicated by the arrows.

\begin{figure}
\includegraphics[width=0.485\textwidth]{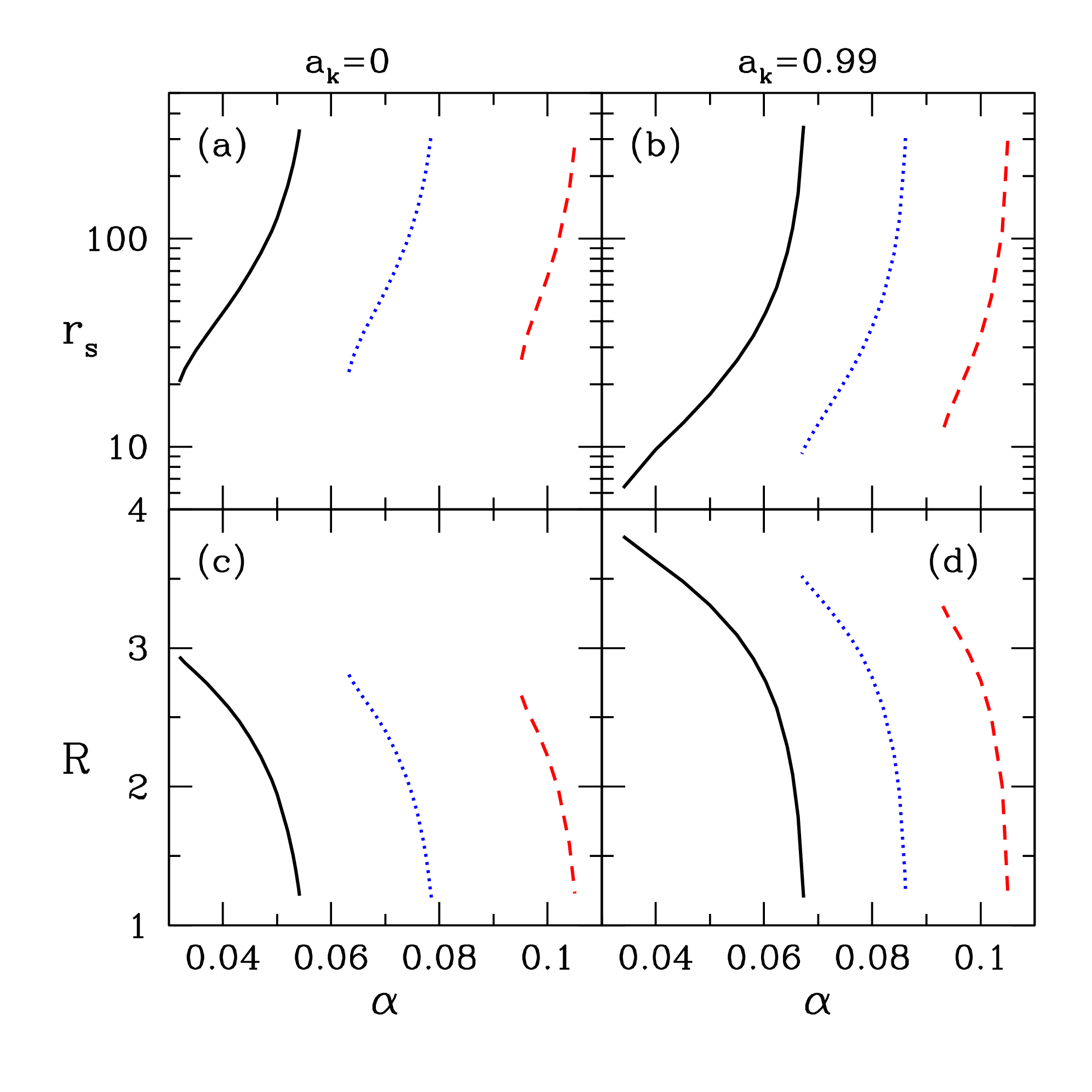}
\caption{Plots of shock location $(r_s)$ and compression ratio $(R)$ as a function of viscosity parameter ($\alpha$). Results corresponding to $a_k=0$ and $a_k=0.99$ are shown in the left and right panels. For $a_{\rm k}=0$, we choose ${\cal E}=1.0002$ and ${\cal L}=2.907$ (solid), $2.737$ (dotted) and $2.567$ (dashed), respectively. Similarly,  for $a_{\rm k}=0.99$, we choose ${\cal E}=1.0002$ and ${\cal L}=1.915$ (solid), $1.865$ (dotted) and $1.815$ (dashed), respectively. See text for details.
	}
\end{figure}

Next, we examine the various shock properties, and the obtained results are depicted in Fig. 4. In the upper panels, we show the variation of shock location as the function of the viscosity parameter ($\alpha$) for flows with ${\cal E} = 1.0002$. In panel (a), we depict the variation of shock location ($r_s$) around the non-rotating black hole (Schwarzschild black hole, $a_k = 0$) where solid, dotted and dashed curves represent the results for ${\cal L} = 2.907$ (black), $2.737$ (blue) and $2.567$ (red), respectively. Similarly, in panel (b), we present the $r_s$ variation for a rotating black hole ($a_k=0.99$) where solid, dotted and dashed curves are obtained for ${\cal L} = 1.915$ (black), $1.865$ (blue) and $1.815$ (red), respectively. In panels (a-b), we observe that for a given ${\cal L}$, shock location recedes away from the black hole as $\alpha$ is increased. This happens because the increase of $\alpha$ enhances the angular momentum transport outwards that boosts the strength of the centrifugal repulsion against the gravity. This eventually compels the shock front to settle down at the larger radii. Interestingly, $\alpha$ can not be increased indefinitely due to the fact that beyond a critical value of viscosity ($\alpha > \alpha^{\rm cri}$), the shock conditions fail to satisfy for a given set of (${\cal L}, {\cal E}$, $a_k$) and therefore, standing shock disappears. However, non-steady shock still may continue to present which we shall discuss in the latter part of this section. Due to the shock transition, the post-shock flow (equivalently PSC) becomes hot and compressed where the swarm of hot electrons are readily available.
These hot electrons eventually reprocess the soft photons from the pre-shock flow via inverse Comptonization process to produce hard radiations. Thus, it is instructive to examine the amount of density compression across the shock front and study its dependencies on $\alpha$. For that, we calculate the compression ratio ($R$), which is defined as the ratio of densities measured immediately after and before the shock transition and is given by $R=\sigma_{+}/\sigma_{-}$. The obtained results corresponding to $r_s$ are depicted in panel (c-d) for $a_k=0$ and $a_k=0.99$, respectively. In both cases, we observe that accreting matter experiences significant compression when shock forms close to the black hole, and the amount of compression gradually decreases with the increase of $\alpha$. Since shock generally forms at smaller radii around the rotating black holes, the overall compression remains higher for $a_k = 0.99$ in comparison with results for $a_k=0$ (see panel c-d).

\begin{figure}
	\includegraphics[width=0.485\textwidth]{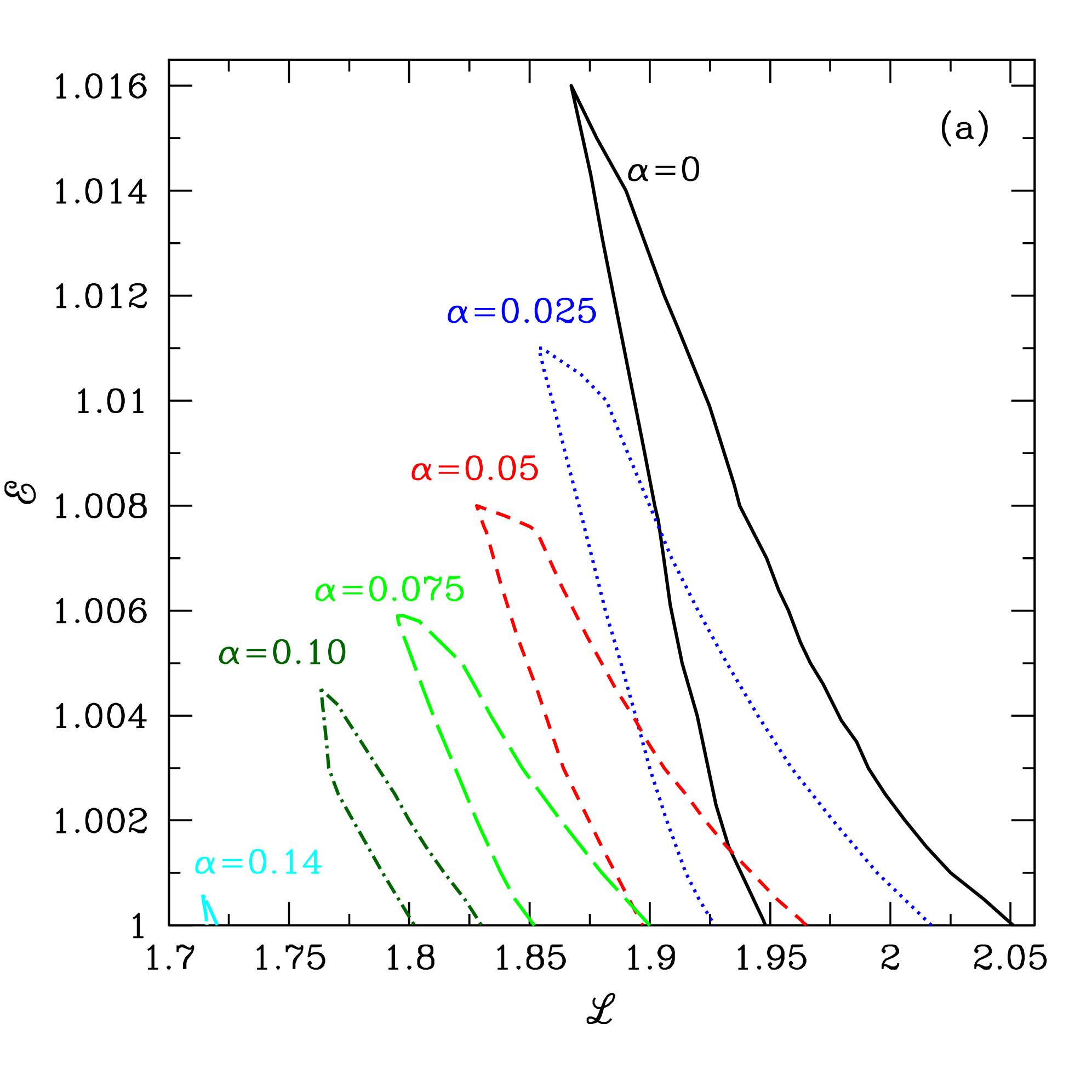}
	\includegraphics[width=0.485\textwidth]{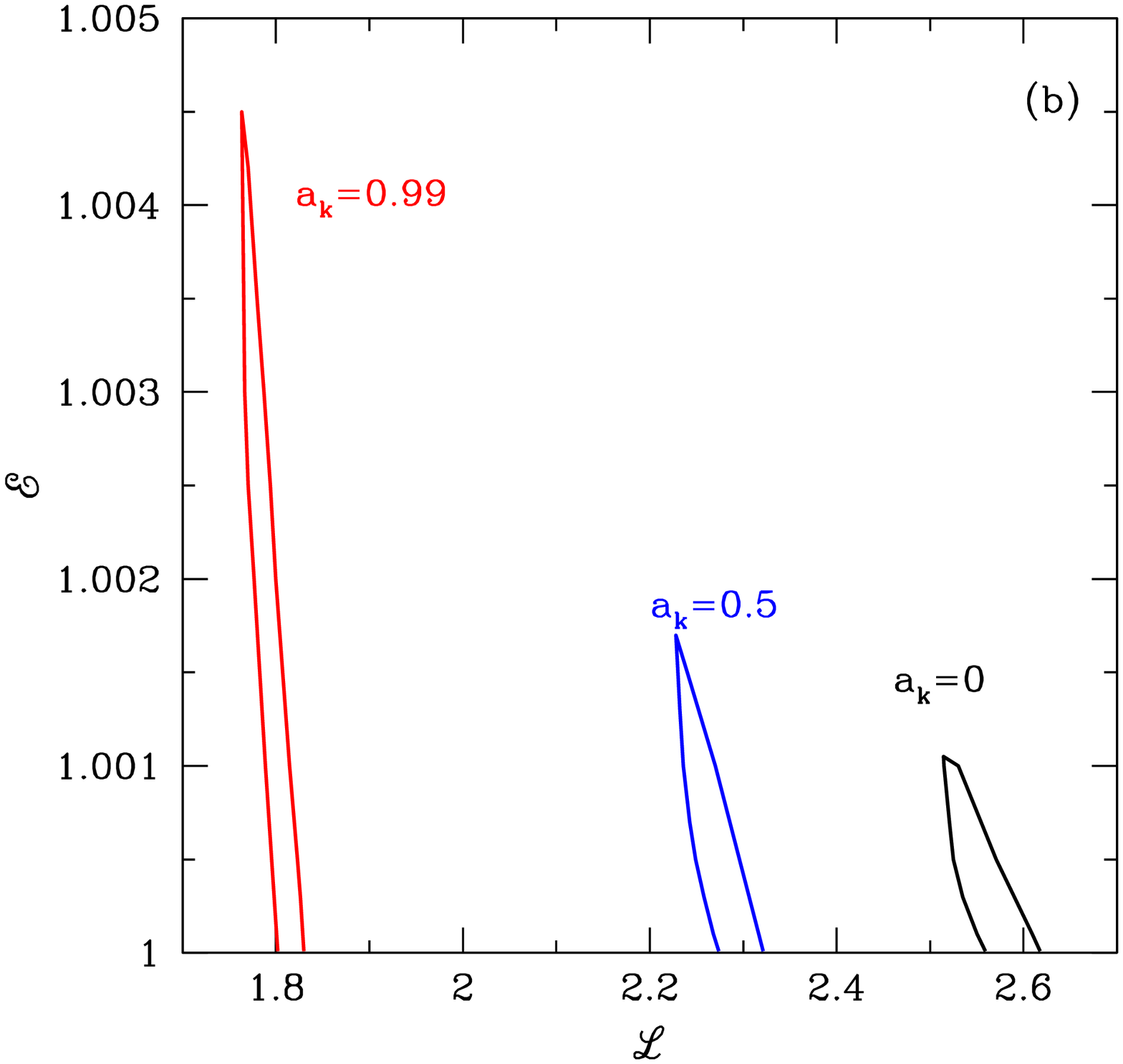}
	\caption{(a) Plots of the shock parameter space in  ${\cal E}-{\cal L}$ plane for different viscosity parameters $(\alpha)$ marked in the figure. Here, we choose $a_{\rm k}=0.99$. The effective domain of the shock parameter space is decreased with $\alpha$ indicating the fact that possibility of shock formation is reduced for dissipative flow. (b) Plots of the ${\cal L}-{\cal E}$ shock parameter space for three different Kerr parameters marked in the figure. Here, $\alpha = 0.1$ is considered. See text for details.
	}
\end{figure}

It is useful to study the range of flow parameters that admit the shock induced global accretion solutions around the Kerr black holes. To do that, in Fig. 5, we identify the effective region of the parameter space in the ${\cal L}-{\cal E}$ plane that provides shock transition. In Fig. 5a, we present the modification of the parameter space with the increase of the viscosity parameter ($\alpha$) for an extremely rotating black hole  $(a_{\rm k}=0.99)$. It is clear from the figure that the wide ranges of ${\cal E}$ and ${\cal L}$ permit shock solutions and the effective region of the parameter space are gradually shrunk as the effect of dissipation, namely, viscosity is increased. In reality, 
when $\alpha$ is increased, it enhances the effect of viscous dissipation in an accretion flow (for example, see the second term in the right hand side of equation (13)) that causes the decrease of global specific energy (${\cal E}$) of the flow. Similarly, ${\cal L}$ is also reduced with the increase of $\alpha$ (see equation (14)). As a result, the overall effective area of the parameter space for shock is decreased with $\alpha$. What is more, is that Fig. 5(a) clearly indicates the existence of the critical viscosity parameter ($\alpha^{\rm cri}$), beyond this value standing shock solution ceases to exist. It is noteworthy that $\alpha^{\rm cri}$ does not possess a universal value. Instead, it depends on the input parameters, namely ${\cal E}$, ${\cal L}$ and $a_k$. In Fig. 5b, we display the classification of the shock parameter space for various $a_k$ values. Here, we fix $\alpha = 0.1$. We find that the accretion flows around the weakly rotating black holes experience shock transitions when ${\cal L}$  
is relatively high compared to the case of rapidly rotating black holes. In reality, accretion flows with angular momentum lower than the marginally stable value are allowed to enter into the black hole. As $a_k$ is increased, the angular momentum at the marginally stable orbit is decreased (Chakrabarti \& Mondal, 2006) and therefore, low angular momentum flows are in general allowed to accrete on to the rapidly rotating black holes. Note that the spin signature in Galactic black holes and AGNs has already been probed observationally \citep[and references therein]{Reynolds-2019}.

\begin{figure} 
	\includegraphics[width=0.485\textwidth]{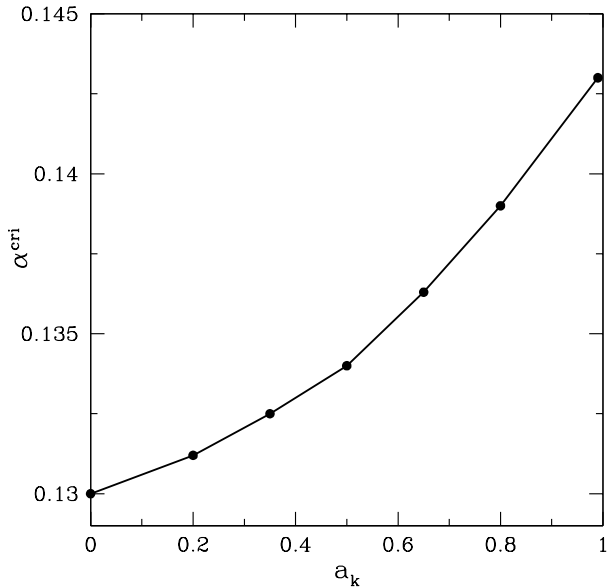}
	\caption{Variation of the critical viscosity parameter $(\alpha^{\rm cri})$ for shock as function of the black hole spin parameter $(a_{\rm k})$. Here, we freely vary the other flow parameters, namely ${\cal E}$ and ${\cal L}$ to compute $\alpha^{\rm cri}$ for a given $a_k$. See text for details.
	}
\end{figure}

In Fig. 6, we show the variation of the critical viscosity parameter ($\alpha^{\rm cri}$) as function of $a_k$. Here, we freely vary both ${\cal L}$ and $\cal E$ and compute $\alpha^{\rm cri}$ for a given $a_k$. By varying $a_k$ in steps, we follow the same procedure to obtain the $\alpha^{\rm cri}$ in the full range of $a_k$ values. In the figure, the chosen values of $a_k$ are denoted by the filled circles which are further joined with the solid lines. The figure clearly indicates that the overall correlation between $\alpha^{\rm cri}$ and $a_k$ remains fairly weak. It may be noted that the obtained $\alpha^{\rm cri}$ are in agreement with the previously reported values from both numerical and observational fronts \citep[and references therein]{Hawley-Krolik01,Hawley-Krolik2002,Penna-etal2013,Martin-etal2019}.

\begin{figure}
	\includegraphics[width=0.485\textwidth]{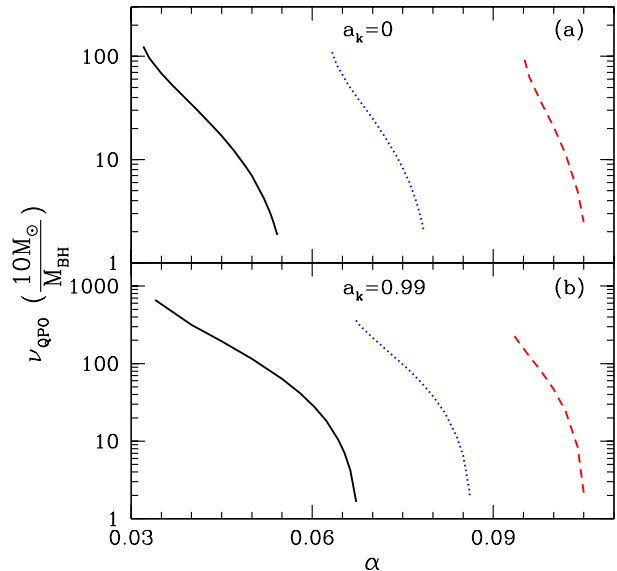}
	\caption{Variation of QPO frequency ($\nu_{QPO}$) 
	as function of viscosity parameter ($\alpha$) for flows with ${\cal E}=1.0002$. In panel (a), we depict the results for $a_{\rm k}=0$ where solid, dotted and dashed curves are for ${\cal L}=2.907$ (black), $2.737$ (blus) and $2.567$ (red), respectively. In panel (b), we show the results corresponding to $a_{\rm k}=0.99$ where solid, dotted and dashed curves are for ${\cal L}=1.915$ (black), $1.865$ (blue) and $1.815$ (red), respectively. See text for more details.
	}
\end{figure}

We further explore the usefulness of the shock wave in an accretion flow. It is already discussed that accretion flow contains standing shock wave provided (i) the flow possesses multiple critical points, (ii) the entropy of the flow at the inner critical point ($r_{\rm in}$) is higher than the outer critical point ($r_{\rm out}$) and (iii) the relativistic shock conditions are satisfied (see equation (24)). However, the situation may arise for an accretion flow where points (i-ii) are fulfilled, but the point (iii) fails to satisfy. In that case, instead of standing shock transition, the shock front demonstrates non-steady behavior. This particularly happens, possibly due to the resonance oscillation where the post-shock cooling time scale remains in accord with the infall time scale of the flow \cite[]{Molteni_etal1996}. 
Indeed, the modulation of the shock front does not remain coherent in general, but yields as Quasi-periodic (QP) in nature. To quantify the frequency of QPOs of the shock front, 
we first calculate the infall time scale of the post-shock flow as $t_{\rm infall}=\int^{r_{\rm H}}_{r_s}dt=\int^{r_{\rm H}}_{r_s}v^{-1}(r)dr$, where $v(r)$ denotes the post-shock velocity and $r_H~(=1+\sqrt{1-a^2_k})$ is the event horizon. Subsequently, we estimate the frequency of the QP oscillation of the shock front ($\nu_{QPO}$) as $\nu_{\rm QPO}=1/t_{QPO} \sim 1/t_{\rm infall}$ 
\cite[and references therein]{Molteni_etal1996,Aktar_etal2015}.
In general, since QPO frequency is expressed in Hertz, $\nu_{\rm QPO}$ is ultimately multiplied with $c^3/GM_{\rm BH}$. For the purpose of representation, we consider the results depicted in Fig. 4 (upper panel) and calculate the corresponding $\nu_{\rm QPO}$. The obtained results are shown in Fig. 7 where the variation of $\nu_{\rm QPO}$ is plotted as function of $\alpha$ for ${\cal E}=1.0002$. In the upper panel (Fig. 7a), we choose $a_k=0$ and the solid, dotted and dashed curves represent the results for ${\cal L}= 2.907$ (black), $2.737$ (blue) and $2.567$ (red). Similarly, in the lower panel (Fig. 7b), we show the results for $a_k=0.99$, where  solid, dotted and dashed curves are for ${\cal L}= 1.915$ (black), $1.865$ (blue) and $1.815$ (red), respectively. We observe that $\nu_{\rm QPO}$ decreases with the increase of the $\alpha$ for both non-rotating as well as rapidly rotating black holes. However, the existence of high frequency QPO (HFQPO) seems to be more viable for the rapidly rotating black holes as shock usually forms closer to the horizon for larger $a_k$ values \cite[][and references therein]{Aktar_etal2017,Dihingia-etal2019}.
 
\begin{figure}
	\includegraphics[width=0.55\textwidth]{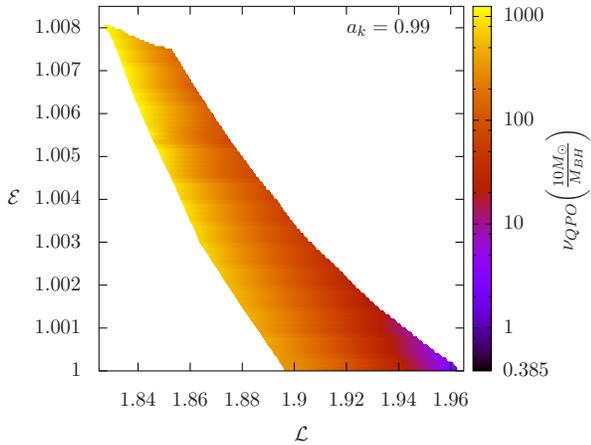}
	\caption{Two dimensional projection of three dimensional plot of 
		$\left[ {\cal L},{\cal E}, \nu_{\rm QPO}\left(\frac{10M_{\odot}}{M_{\rm BH}}\right) \right]$ for $a_{\rm k}=0.99$ and $\alpha=0.05$. Color coded bar indicates the frequency range in logarithmic scale. See text for details.
	}
\end{figure}

In Fig. 8, we redraw the shock parameter space for $a_k=0.99$ and $\alpha = 0.05$, where the two-dimensional projection of the three-dimensional plot spanned with ${\cal L}$, ${\cal E}$ and $\nu_{QPO}$ is depicted. In the right side of the figure, the color coded vertical bar indicates the range of $\nu_{QPO}$ (in units of $10M_\odot/M_{BH}$) obtained by using the post-shock velocity profile as described above. We find that for a given ${\cal E}$, accretion flow generally exhibits high frequency QPOs provided ${\cal L}$ is relatively small and {\it vice versa}. These findings are in agreement with the results of Fig. 7, as shocks generally settle down at smaller radii for flows with lower ${\cal L}$ (see Fig. 4) that yield high $\nu_{QPO}$ values. Also, we observe that Fig. 8 encompasses the QPO frequencies starting from milli-Hz ($\sim 0.386$ Hz) to kilo-Hz ($\sim 1312$ Hz) range for a $10M_\odot$ black hole. This evidently indicates that the present formalism would be capable of rendering the QPO frequencies observed from the Galactic black hole sources \cite[and references therein]{Remillard-etal1999,Strohmayer2001,Belloni_etal2002,Belloni_etal2005,Remillard_McClintock2006,Altamirano-Belloni2012,Belloni-etal2012,Nandi_etal2012,Belloni-Altamirano2013,Iyer_etal2015,Sreehari-etal2019b}.

\begin{figure}
	\includegraphics[width=0.5\textwidth]{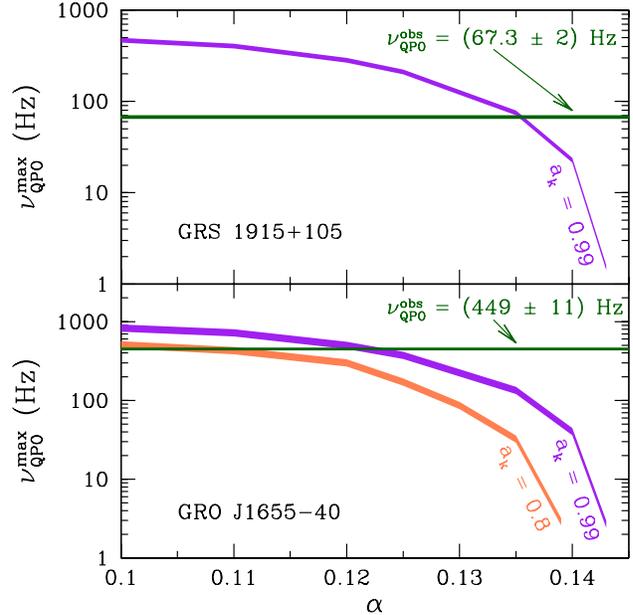}
	\caption{Comparison of theoretically calculated maximum QPO frequency ($\nu^{\rm max}_{QPO}$) with observed HFQPOs. In the upper panel, we present the results for GRS $1915 + 105$, where shaded region (in purple) are obtained for $a_k=0.99$ and $M_{BH}=10.1\pm0.6~M_\odot$. The horizontal thick line (green) indicates the observed HFQPO as $\nu^{\rm obs}_{QPO}\sim 67.3\pm2$ Hz. In the lower panel, results corresponding to GRO J$1655 - 40$ is shown. Here, shaded regions are for $a_k=0.99$ (purple) and $a_k=0.8$ (orange), respectively with $M_{BH}=5.1-6.3~ M_\odot$. The horizontal thick line (green) indicates the observed HFQPO as $\nu^{\rm obs}_{QPO}\sim 449\pm11$ Hz. See text for details.
	}
\end{figure}

At the end, we compare theoretically obtained maximum QPO frequency ($\nu^{\rm max}_{QPO}$) with observed HFQPOs. For that, we choose two well studied Galactic black hole sources, namely GRS $1915 + 105$ and GRO J$1655 - 40$ as they are known to exhibit HFQPOs. 
For GRS $1915 + 105$, the mass and spin are well constrained and this source exhibits HFQPO as $\nu^{\rm obs}_{QPO}\sim 67.3\pm2$ Hz \cite[]{Morgan-etal1997,Belloni-Altamirano2013}. 
In this work, we consider $M_{BH}=10.1\pm0.6~M_\odot$ \cite[]{Steeghs-etal2013} and $a_k=0.99$ \cite[]{Miller-etal2013}. Using these fundamental parameters, we calculate the maximum QPO frequency ($\nu^{\rm max}_{QPO}$) as function of viscosity parameter ($\alpha$). This result is shown in the upper panel of Fig. 9, where the shaded region (in purple) is obtained considering the uncertainty in the estimate of the source mass. The figure indicates that $\nu^{\rm obs}_{QPO} < \nu^{\rm max}_{QPO}$ for $\alpha \lesssim 0.135$.  
For GRO J$1655 - 40$, the source mass is estimated using the dynamical method as $M_{BH}=5.1-6.3~M_\odot$ \cite[]{Greene-etal2001,Beer-etal2002}.
Interestingly, contradictory claims appear in the measurement of the spin parameter \cite[]{Abramowicz-Kluzniak2001,Shafee-etal2006,Motta-etal2014,Aktar_etal2017,Dihingia-etal2019} which is yet to be settled. 
Hence, in this work, we consider $0.8 \le a_k \le 0.99$ for representation and calculate $\nu^{\rm max}_{QPO}$ as before. The obtained results are depicted in the lower panel of Fig. 9, where $a_k$ values are marked and shaded regions (in purple and orange) are because of the mass range of GRO J$1655 - 40$. We observe that for this source, when $a_k=0.99$ (rapidly rotating), $\nu^{\rm obs}_{QPO} < \nu^{\rm max}_{QPO}$ with $\alpha \lesssim 0.125$, whereas for $a_k=0.8$ (moderately rotating), $\nu^{\rm obs}_{QPO} < \nu^{\rm max}_{QPO}$ with $\alpha \lesssim 0.1$.

It may be noted that the estimated ranges of $\alpha$ for the above sources (GRS $1915 + 105$ and GRO J$1655 - 40$) fairly are in agreement with the results of numerical simulations \cite[]{Hawley-Krolik01,Hawley-Krolik2002,Penna-etal2013}.

\section{Conclusions}

In this work, we study the relativistic viscous accretion flow around a Kerr black hole considering  the general relativistic approach. We compose the governing equations that describe the motion of the accreting matter in an accretion disc and employing the boundary parameters, we solve these equations to obtain the transonic accretion solutions. We find that depending on the flow parameters, accretion flows may contain multiple critical points and such flows are astrophysically important as they may harbor shock waves provided the relativistic shock conditions are favorable. Below, we summarize our findings based on the present work.

(1) We study the transonic properties of the accretion flow and find that 
flows continue to possess more than one critical point when the input parameters are chosen appropriately. We obtain all the possible accretion solutions around the Kerr black holes by tuning the flow parameters and separate the parameter space in the ${\cal L}-{\cal E}$ plane according to the nature of the accretion solutions (see Fig. 2). It may be noted that we find a new-type of accretion solution ($A^\prime$ in Fig. 2).

(2) To the best of our knowledge, for the first time, we show that relativistic viscous accretion solutions around the Kerr black holes experience discontinuous transitions in the flow variables in the form of shock waves
provided the relativistic shock conditions are satisfied (see Fig. 3). We study the properties of shock waves, namely shock location ($r_s$) and compression ratio ($R$) and examine their dependencies on the input parameters. We observe that shock fronts in general settle down to the smaller radii when the flows accrete on to the rapidly rotating black holes and {\it vice-versa}. As $r_s$ is small, flow experiences more compression at the shock discontinuity resulting in high values of $R$ (see Fig. 4). 

(3) We make an effort to constrain the range of the input parameters that admit shock induced global accretion solutions around the Kerr black holes. We find that standing shocks are permitted for the wide ranges of the ${\cal L}$ and ${\cal E}$ for inviscid flow. As the viscosity is increased, the ranges of parameters are gradually shrunk in the lower ${\cal L}$ and ${\cal E}$ domain, and ultimately standing shock disappears when the viscosity exceeds its critical limit ($\alpha^{\rm cri}$) (see Fig. 5). We calculate $\alpha^{\rm cri}$ by freely varying the input parameters and observe that $\alpha^{\rm cri}$ varies with $a_k$ weakly (see Fig. 6).

(4) It is intriguing to note that accretion flow often initiates the 
modulation of its inner part, particularly when the relativistic shock conditions are not favorable, but the entropy of the inner critical point ($r_{\rm in}$) is higher than the outer critical point ($r_{\rm out}$).  
The outcome of this effect results the QPO of emitted radiations that are commonly observed from the Galactic black hole sources during their different evolutionary phases.
Using a phenomenological approach, we estimate the frequency of the QPO ($\nu_{QPO}$) and study the role of the input parameters on $\nu_{QPO}$. With this, we identify the shock parameter space in ${\cal L}-{\cal E}$ plane in terms of the $\nu_{QPO}$ and observe that the present formalism is capable of explaining the QPOs in the frequency range starting from milli-Hertz to kilo-Hertz (see Fig. 8). Using our model formalism, we phenomenologically estimate the ranges of viscosity parameter as $\alpha\lesssim 0.135$ and $0.125$ for GRS $1915 + 105$ and GRO J$1655 - 40$, respectively that could possibly account for the observed HFQPOs (Fig. 9). It may be noted that the above findings are purely indicative. For the quantitative estimates of the range of viscosity parameters, time-dependent numerical modeling is required which is beyond the scope of the present paper.

Finally, we point out that in this paper, we have imposed several assumptions and approximations. For example, we neglect the radiative cooling processes, namely bremsstrahlung, synchrotron, and Compton coolings mechanisms although their presence are inevitable in the accretion disc. Furthermore, we assume that strong coupling exists between ion and electron that renders the flow to maintain a single temperature all throughout the disc. However, because of the weak ion-electron coupling, the two-temperature flow structure seems to be viable at least in the inner part of the disc. Moreover, we ignore the mass loss from the disc as well. Although the implementation of all these aspects is beyond the scope of the present work, however, we plan to incorporate them in our future works. 

\section*{Acknowledgments}

Authors thank the anonymous reviewer for constructive comments and suggestions that help to improve the quality of the paper.
AN thanks GD, SAG; DD, PDMSA and Director, URSC for encouragement and continuous support to carry out this research. ID thanks Bibhas Ranjan Mahji, Rashidul Islam, Purusottam Ghosh and Pankaj Saha for fruitful discussions.

\appendix
\bsp   
\section{Explicit expression of viscous stress}
The components of viscous stresses $\sigma^r_\phi$ and $\sigma^r_t$ in co-rotating frame are expressed as,
$$
2\sigma^r_\phi={\cal A}_1 + {\cal A}_2\frac{dv}{dr}+{\cal A}_3\frac{d\lambda}{dr},
\eqno(A1)
$$
$$
2\sigma^r_t={\cal B}_1 + {\cal B}_2\frac{dv}{dr}+{\cal B}_3\frac{d\lambda}{dr},
\eqno(A2)
$$
where
$$\begin{aligned}
{\cal A}_1&=\frac{\left({\cal A}_{11}+{\cal A}_{12}+ {\cal A}_{13}\right)\gamma_v^3}{{\cal S}},\\
{\cal A}_2&=\frac{4 \Delta  \lambda  v \gamma_v^5}{3 r^2 \sqrt{{\cal PQ}} },\\
{\cal A}_3&=\frac{\Delta \gamma_v^3 }{r^2  {\cal P}^{3/2}\sqrt{{\cal Q}}},\\
{\cal B}_1&=\frac{\left({\cal B}_{11}+{\cal B}_{12}\right)\gamma_v^3}{{\cal S}},\\
{\cal B}_2&=-\frac{4 \Delta v \gamma_v^5}{3r^2  \sqrt{PQ}},~~{\rm and}\\
{\cal B}_3&=-\frac{(2a_{\rm k}+(r-2) \lambda )\gamma_v^3 }{r^3 \left({\cal PQ}\right)^{3/2}}.\\
\end{aligned}$$
Here, we write
$$\begin{aligned}
{\cal A}_{11}&=-3a_{\rm k} r \left(a_{\rm k}^2+3 r^2\right) \left(a_{\rm k}^2 (r+2)+r^3\right),\\
{\cal A}_{12}&=r v^2 \left(3a_{\rm k} \left(a_{\rm k}^2 
+ 3 r^2\right)-2 \left(a_{\rm k}^2 (r+3)-(r-3) r^2\right) \lambda \right) \\
&\times \left(a_{\rm k}^2 (r+2)-\lambda  (4a_{\rm k}+(r-2) \lambda )+r^3\right),\\
{\cal A}_{13}&=-6  a_{\rm k} \lambda^2  \left(a_{\rm k}^2 (r+1) (r+4)-a_{\rm k} (r+2) \lambda +4 r^3\right)\\
&+6 \lambda\left(a_{\rm k}^2 (r+2)+r^3\right) \left(a_{\rm k}^2 (2 r+1)-(r-3) r^3\right),\\
{\cal B}_{11}&=2 r v^2 \left((r-3) \left(a_{\rm k}^2+2 r^2\right)+3a_{\rm k} \lambda \right)\\
&\times \left(a_{\rm k}^2 (r+2)-\lambda  (4a_{\rm k}+(r-2) \lambda )+r^3\right)\\
{\cal B}_{12}&=3 (2a_{\rm k}+(r-2) \lambda ) \\
&\times \left(a_{\rm k}^3 (r-2)+2 \lambda  \left(a_{\rm k}^2 (1-2 r) 
+ a_{\rm k} r \lambda +(r-3) r^3\right)+5a_{\rm k} r^3\right)\\
{\cal P}&=\frac{a_{\rm k}^2 (r+2)-\lambda  (4a_{\rm k}+(r-2) \lambda )+r^3}{a_{\rm k}^2 (r+2)-2a_{\rm k} \lambda +r^3},\\
{\cal Q}&=\frac{a_{\rm k}^2 (r+2)-2a_{\rm k} \lambda +r^3}{r \Delta},~~{\rm and}\\
{\cal S}&=3r^6 \Delta \left({\cal PQ}\right)^{3/2},
\end{aligned}$$
where quantities have their usual meanings.

\section{Detail Expressions of the governing  equations}
The radial momentum, angular momentum, and energy equations are obtained as,
$$
{\cal R}_0+{\cal R}_1\frac{dv}{dr}+{\cal R}_2\frac{d\Theta}{dr}+{\cal R}_3\frac{d\lambda}{dr}=0,
\eqno(A3)
$$
$$
{\cal E}_0+{\cal E}_1\frac{dv}{dr}+{\cal E}_2\frac{d\Theta}{dr}+{\cal E}_3\frac{d\lambda}{dr}=0, ~~{\rm and}
\eqno(A4)
$$
$$
{\cal L}_0+{\cal L}_1\frac{dv}{dr}+{\cal L}_2\frac{d\Theta}{dr}+{\cal L}_3\frac{d\lambda}{dr}=0,
\eqno(A5)
$$
where
$$\begin{aligned}
{\cal R}_0&=\left(\frac{\partial\Phi}{\partial r}\right)_\lambda+\frac{\Theta}{f+2\Theta}\bigg[\frac{2-2 r}{a_{\rm k}^2+(r-2) r}+\frac{{\cal F}_1'}{{\cal F}_1}\\
&-\frac{3}{r}-\frac{2 \lambda  \left(a_{\rm k}^3+\lambda  \left(-2a_{\rm k}^2 + a_{\rm k} \lambda +(r-3) r^2\right)+3a_{\rm k} r^2\right)}{{\cal R}_{\rm D}}\bigg],\\
{\cal R}_1&=v\gamma_v^2 - \frac{2 \Theta\gamma_v^2}{v (f+2 \Theta )},\\
{\cal R}_2&=\frac{1}{f+2\Theta},\\
{\cal R}_3&=\frac{2 \Theta\left\{ (r-2) \lambda  \left(a_{\rm k}^2 (r+2) 
- a_{\rm k} \lambda +r^3\right) + a_{\rm k} \left(a_{\rm k}^2 (r+2)+r^3\right)\right\} }{(f+2 \Theta){\cal R}_{\rm D} },\\
{\cal L}_0&=\frac{\gamma_vh \lambda {\cal L}_{\rm N}\sqrt{{\cal Q}}}{\left(a_{\rm k}^2 (r+2)-2a_{\rm k} \lambda +r^3\right)^2 {\cal P}^{3/2}} +\frac{\partial\left(\Lambda {\cal A}_1\right)}{\partial r},\\
{\cal L}_1&=\frac{h \lambda  v \gamma_v^3}{ \sqrt{{\cal PQ}}} + \frac{\partial\left(\Lambda {\cal A}_1\right)}{\partial v},\\
{\cal L}_2&=\frac{2\gamma_v\lambda (N+1)}{\tau \sqrt{{\cal PQ}}} + \frac{\partial\left(\Lambda {\cal A}_1\right)}{\partial \Theta},\\
{\cal L}_3&=\frac{h \gamma_v}{{\cal P}^{3/2}\sqrt{{\cal Q}}}+\frac{\partial\left(\Lambda {\cal A}_1\right)}{\partial \lambda},\\
{\cal E}_0&=\frac{h\gamma_v\sqrt{Q}}{\left(a_{\rm k}^2 (r+2)-2a_{\rm k} \lambda +r^3\right)^2{\cal P}^{3/2}} + \frac{\partial\left(\Lambda {\cal B}_1\right)}{\partial r},\\
{\cal E}_1&=\frac{hv\gamma_v^3}{\sqrt{{\cal PQ}}} + \frac{\partial\left(\Lambda {\cal B}_1\right)}{\partial v},\\
{\cal E}_2&=\frac{2(N+1)\gamma_v}{\tau\sqrt{PQ}} + \frac{\partial\left(\Lambda {\cal B}_1\right)}{\partial \Theta},~~{\rm and}\\
{\cal E}_3&=\frac{h (2a_{\rm k}+(r-2) \lambda )\gamma_v}{r \Delta{\cal P}^{3/2}{\cal Q}^{3/2}} + \frac{\partial\left(\Lambda {\cal B}_1\right)}{\partial \lambda}.\\
\end{aligned}$$

\noindent Here, we write
$$\begin{aligned}
{\cal R}_{\rm D}&=\left(a_{\rm k}^2 (r+2)-2a_{\rm k} \lambda +r^3\right)\\
\times& \left(a_{\rm k}^2 (r+2)-\lambda  (4a_{\rm k}+(r-2) \lambda )+r^3\right)\\
{\cal L}_{\rm N}&=a_{\rm k}^4-2a_{\rm k} \left(a_{\rm k}^2+r (3 r-4)\right) \lambda +\left(a_{\rm k}^2-(r-2)^2 r\right) \lambda ^2\\
&+2a_{\rm k}^2 (r-2) r+r^4,\\
\Phi&=\frac{1}{2}\ln\left[\frac{r\Delta}{a_{\rm k}^2 (r+2)-4 a_{\rm k} \lambda +r^3-\lambda ^2 (r-2)}\right],\\
{\cal F}_1&=\frac{\left(a_{\rm k}^2+r^2\right)^2+2a_{\rm k}^2 \Delta }{\left(a_{\rm k}^2+r^2\right)^2-2a_{\rm k}^2 \Delta },~~{\rm and}\\
\Lambda&=\frac{\alpha a_{\rm s} H r}{v\gamma_v\sqrt{\Delta}},
\end{aligned}$$
where quantities have their usual meanings.

\section{Explicit expression of Wind equation}
We obtain the wind equation which is given by,
$$
\frac{dv}{dr}=\frac{{\cal N}\left(r,v,\lambda,\Theta\right)}{{\cal D}\left(r,v,\lambda,\Theta\right)},
\eqno(A6)
$$
where
$$
\begin{aligned}
{\cal N}\left(r,v,\lambda,\Theta\right)&={\cal L}_3 \left({\cal E}_2 {\cal R}_0-{\cal E}_0 {\cal R}_2\right)
+{\cal L}_2 \left({\cal E}_0 {\cal R}_3-{\cal E}_3 {\cal R}_0\right)\\
&+{\cal L}_0 \left({\cal E}_3 {\cal R}_2-{\cal E}_2 {\cal R}_3\right)~~{\rm and}\\
{\cal D}\left(r,v,\lambda,\Theta\right)&={\cal L}_3 \left({\cal E}_1 {\cal R}_2-{\cal E}_2 {\cal R}_1\right)
+{\cal L}_1 \left({\cal E}_2 {\cal R}_3-{\cal E}_3 {\cal R}_2\right)\\
&+{\cal L}_2 \left({\cal E}_3 {\cal R}_1-{\cal E}_1 {\cal R}_3\right).\\
\end{aligned}
$$

\end{document}